\documentclass{tufte-handout}

\title{Breaking the Barriers to True Augmented Reality}
\author{Christian Sandor$^{\dagger}$, Martin Fuchs$^{\star}$, Alvaro Cassinelli$^{\dagger}$, Hao Li$^{\circ}$, Richard Newcombe$^{\times}$, Goshiro Yamamoto$^{\dagger}$, Steven Feiner$^{\wedge}$\newline
$^{\dagger}$Nara Institute of Science and Technology, Graduate School of Information Science, Japan\newline
${^\star}$Universit\"at Stuttgart, Visualization Research Center, Germany\newline
${^\circ}$University of Southern California, Department of Computer Science, USA\newline
${^\times}$University of Washington, Department of Computer Science and Engineering, USA\newline
${^\wedge}$Columbia University, Department of Computer Science, USA}

\usepackage{mdwlist} 
\usepackage{graphicx} 
  \setkeys{Gin}{width=\linewidth,totalheight=\textheight,keepaspectratio}
  \graphicspath{{graphics/}} 
\usepackage{amsmath}  
\usepackage{booktabs} 
\usepackage{units}    
\usepackage{multicol} 
\usepackage{lipsum}   
\usepackage{fancyvrb} 
  \fvset{fontsize=\normalsize}

\begin{document}
\begin{fullwidth}
\maketitle
\end{fullwidth}

\newthought{In recent years, Augmented Reality (AR) and Virtual Reality (VR)} have gained considerable commercial traction, with Facebook acquiring Oculus VR for \$2 billion, Magic Leap attracting more than \$500 million of funding, and Microsoft announcing their HoloLens head-worn computer. Where is humanity headed: a brave new dystopia---or a paradise come true?

In this article, we present discussions, which started at the symposium ``Making Augmented Reality Real'', held at Nara Institute of Science and Technology in August 2014. Ten scientists were invited to this three-day event, which started with a full day of public presentations and panel discussions (video recordings are available at the event web page\cite{marr-web}), followed by two days of roundtable discussions addressing the future of AR and VR.

Both AR and VR facilitate experiences that can go far beyond what we encounter in our everyday lives. VR completely immerses users; for example, we could feel as if we were walking on the moon. On the other hand, AR embeds content directly into the user's real environment; for example, we could preview a piece of furniture in our actual living room before we buy it. We believe that the current rapid convergence of enabling technologies will lead to AR and VR experiences that are barely distinguishable from reality.

From a technical perspective, AR is clearly more challenging. We not only have to create highly responsive and realistic virtual media, but we have to additionally fuse them coherently with the user's real environment. In this article, we focus on AR, as any AR system can be trivially turned into a VR system---we just need to block the user's experience of the real world. To distinguish our vision of indistinguishably real AR from currently available AR systems, we call the former True AR.

In the remainder of this article, we first discuss high-level aspects of True AR: Why is it desirable (Section 1)? What exactly do we mean by True AR? Is it possible to define a clear-cut test for it (Section 2)? What are the approaches to realizing True AR (Section 3)? Then, we discuss challenges in more detail: technical (Section 4) and ethical (Section 5) and draw overall conclusions (Section 6).

\vspace{1em}
\section{True AR: Why?}
\vspace{-1em}
\newthought{After our needs for survival are fulfilled}, we are driven by the desire for experiences. Actual reality, however, confronts us with limitations.  These limitations can be physical (we will never move faster than light), biological (humans cannot breathe on their own under water), ethical (we should not infringe on the freedom of others in pursuing our own), or simply practical---while humans have walked on the moon before, the sheer difficulty and expense of doing that will keep this actual experience from most of us forever.

Our imagination is not subject to these limitations, and we can overcome any boundary. For a long time, humankind has supported imagination through the creation and consumption of media, starting with stories told around campfires and ending with the ubiquitous presence of television today. Still, conventional media deliver only a subset of the stimuli possible within actual reality.

With the advent of VR technology, we can create media experiences far beyond the passive consumption of a prerecorded television show or movie. Within digitally simulated environments, we can interact with a responsive virtual world. AR even surpasses VR in interactivity: it not only reacts to the observer, but can also incorporate the observers' surroundings. While AR is technically a proper superset of VR, its utility begins with much lower demands on perceived realism than VR. For example, a heads-up display in a car that reliably highlights pedestrians at night as silhouettes needs no photorealistic simulation of how light interacts with the world to be useful.

As technology continues to evolve, the boundaries between traditional media (e.g., phone calls vs. interactive video, or newspapers vs. interactive web pages) become diffuse.  Similarly, the boundary between passive media and interactive VR and AR begin to blur, with the first interactive movies being developed for VR headsets. Therefore, we believe that increasing demand will drive True AR development to the point that we will experience it in our lifetimes. But, are there clear criteria by which we can judge the achievement of True AR?

\section{True AR: What?}
\vspace{-1em}
\newthought{In 1950, Alan Turing introduced the Turing Test}, an essential concept in the philosophy of Artificial Intelligence (AI). He proposed an ``imitation game'' to test the sophistication of AI software. At its core is a precise test protocol through which human participants are to determine whether a conversation partner is an actual human or an AI simulation. Similar tests have been suggested for fields including Computer Graphics\cite{gfx-turing} and Visual Computing\cite{vc-turing}. In this section, we discuss an application of this idea of perfection as indistinguishability, to yield an AR Turing Test.

A charm of the original Turing Test is that it restricts possible user interaction with the simulation so that the most essential qualities of the simulation are tested (cognition, not the graphical rendering of a human being, which in the 1950s would have been far beyond the capabilities of any existing computer). Similarly, we believe that it makes little sense to define a single true-for-all-time test protocol, as the applications of AR vary too widely. Instead, we focus on the dimensions along which the interactions in a useful test scenario can be restricted.
Informally speaking, we define True AR to be a modification of the user's perception of their surroundings that cannot be detected by the user. The most obvious parameter of the test protocol then should be which senses can be used: even the most sophisticated visual display will immediately fail a test in which users can use their hands to touch objects in order to tell the real from the virtual.

A second dimension comprises the consistency of the simulated world with actual reality: Can users use a (potentially) virtual microscope to examine a surface? How about a CT scanner? Can they cook a meal and eat it?  Or, spend sufficient time to see whether the change of seasons is consistent with their expectations?

Third, how is the transition between actual and augmented reality handled, if special devices need to be worn?  One possibility would be the simultaneous exposure to virtual and actual reality to be compared:  have the subject put on all necessary displays, then remove a box that occluded the view. Which of the two apples the subject sees is real, which a simulation?

\section{True AR: How?}
\vspace{-1em}
\newthought{In the past, several potential approaches} for True AR have been discussed. Two of these approaches are presented in early work by Ivan Sutherland in the 1960s. Describing a hypothetical ``Ultimate Display'', he wrote\cite{theultimatedisplay}:

\begin{quote}
The ultimate display would, of course, be a room within which the computer can control the existence of matter. A chair displayed in such a room would be good enough to sit in. Handcuffs displayed in such a room would be confining, and a bullet displayed in such a room would be fatal.
\end{quote}

Here, he defined the ultimate vision of True AR: creating, manipulating, and destroying atoms, in the spirit of the Star Trek Holodeck and earlier science fiction stories. 

However, there are huge technical challenges involved with rapidly altering atoms, as would be required to implement this scenario. Sutherland tackled a subset of this problem three years later\cite{sutherland1968}. He developed a stereoscopic head-worn display, whose 3D position and orientation were tracked. This allowed him to display objects that appeared to exist within the user's physical environment. Instead of manipulating the atoms surrounding the user, the user's perception of the environment is carefully measured and an illusion of those manipulated atoms is displayed.

In the remainder of this section, we present the four main approaches for achieving True AR, sorted in decreasing order of physicality: from ``manipulating atoms'' to ``manipulating perception'':
\begin{itemize*}
\item Controlled Matter
\item Surround AR
\item Personalized AR
\item Implanted AR
\end{itemize*}

\newthought{Controlled Matter.} As Sutherland suggests, manipulating atoms is in some sense the ideal way to implement True AR. Creating physical objects by rapidly reconfiguring atoms would create physical objects consistent over all interaction modalities and all users. However, there are huge technical hurdles for this, including energy and safety requirements. Although there is no general-purpose display in sight, several displays for special cases have been developed.

For example, displays that use magnetic fields to rapidly create shapes out of ferromagnetic fluid have been developed\cite[-4cm]{kodama}. A limitation of this approach is that the resulting shapes are not solid and cannot be touched due to the safety concerns when interacting with strong magnetic fields. The solidity limitation is addressed by another class of displays that levitate solid objects in a field of overlaid ultrasonic\cite[-4.5cm]{ochiai} or magnetic waves \cite[-2.9cm]{lee}; again, scaling these up raises safety because of the required strength of overlaid fields.  Another popular approach is to use tabletop pin-arrays\cite[-1.4cm]{ikei} that are solid and can be touched. However, actuated pins severely limit the kinds of shapes that can be displayed to a height field, which precludes overhangs and multiple layers.

\newthought{Surround AR.} The next best thing to manipulating atoms is manipulating photons that reach the user. Although the user cannot touch objects created like this, they can be potentially made visually indistinguishable from physical reality; since the majority of our perceptual bandwidth is allocated to the visual sense, visual augmentations are extremely important. 

For maximum visual fidelity, the user could be surrounded with displays that recreate the wave front of photons that would be emitted from the environment to be presented---or, at least, a distribution of light that has the same visual effect on a human observer. Such light field displays would provide several advantages: no tracking of users would be required; hence, they would automatically support multiple users in the same space. Even before the digital revolution, integral photography\cite{lippman} approached this effect by means of lenslet arrays affixed to photographic plates. Holography\cite{gabor} stores wave fronts as interference patterns. The practical limits on achievable resolution and the immense amount of data required for large, dynamic displays have kept this approach back so far, though, and the next approach has been used more widely. For haptics, a promising direction could be stimulating the user's skin through ultrasound waves, as demonstrated by Long and colleagues\cite{long}. 

\newthought{Personalized AR.} Instead of displaying the full environment around the user, we could display the subset that the user is currently observing. Such a targeted manipulation of the user's perception is much more feasible. Sutherland's head-worn display is the first embodiment of this approach.  The downside of this approach is that we need to track the user's viewpoint at a sufficiently high update rate and low latency. In addition, if multiple users are involved, the effort is multiplied and their displays must be synchronized; recent examples include Meta, HoloLens, and Lumus. 

\newthought{Implanted AR.} In our progression of virtualization of displays, the extreme is not to send any photons towards the user; instead, we directly manipulate the user's perceptual system. This approach has a long history of being depicted in science fiction, including movies such as The Matrix and Total Recall. Futurists have predicted that this is the ultimate way to change our realities; for example\cite{kurzweil2000age}:``nanotechnology will augment our bodies . . . as humans connect to computers via direct neural interfaces or live full-time in virtual reality.'' Recently, this approach has gained increasing attention, based on the success stories around the bionic eye.  By directly stimulating the retina or the visual cortex with electricity, blind persons can regain vision. For example, the Argus system has been approved by the European Union for commercial use in 2011 and by the FDA in 2013. So far, this approach has not been applied to augment a healthy person's reality.

In summary, we think that the most promising approach in the short term is Personalized AR. Although the other approaches are promising in the long term, significant challenges remain to be solved. 

\section{Facing Reality: Technical Barriers} 
In exploring the limits to True AR, we can now turn our attention to technical challenges. As the visual channel has the highest bandwidth, it has historically received most of our attention---most universities feature visual computing in one aspect or another as part of a regular computer science curriculum, few address acoustics, for example. Other modalities are even less developed (touch, smell, taste). In following this tradition, we will focus on visual interaction as well, but consider it as exemplary for the challenges posed by the other modalities.

A prerequisite for achieving True AR is the ability to photorealistically capture reality and display it to the user, possibly after altering it. In this section, we first discuss display (Section 4.1); we argue that the only way to achieve True AR is to implement a light field display. Next, we discuss capture, with a special focus on capturing light fields (Section 4.2).

\subsection{The Case for Light Field Displays}

For the visual channel, technological progress has been somewhat uneven. Color reproduction is, while not a completely solved problem, already rather satisfactory, and the available solution space is rather well understood.

Also, the spatial resolution of modern displays approximates what the human eye can reasonably discern; so called retina displays are an example, and in the pixels-per-inch (ppi), race some manufacturers even exceed it. One important aspect is missing, however: expressing human visual system (HVS) capabilities via ppi are relative to a flat device.  In interacting with the world, we need something better.  And indeed, the HVS can be better modeled with discernable-resolution-per-angle-seen, and ppi-at-distance can be derived from that; however, if we only recreate that, we treat the eye as a pinhole camera, which is a fundamentally incomplete model.

The eye is a complex optical system; specifically, it contains a lens of variable refractive power and a variable aperture. This system adapts, and in addition to the image we see, it gives us unconscious feedback on its optical state. Additionally, the optics in the eye change as we pass them over the scene in order to achieve vergence and accommodation, and this generates visual impressions that we expect and miss in flat images (for instance, changing bokeh or out-of-focus blur). In order to achieve True AR, we need to present each eye with a light distribution equivalent to the light leaving a real scene.

The distribution of light in space is exhaustively modeled with the {\bf plenoptic function}\cite{Adelson91theplenoptic}. It is at least six-dimensional: three dimensions for the observer's viewpoint, two for the viewing direction, and one for the wave length; its dense sampling and reconstruction is expensive. One assumption that makes either task easier is that the environment is observed---or rendered---from the outside, thus keeping spectral radiance the same along straight lines and effectively removing one dimension from the viewpoint, reducing the representation to a {\bf light field}\cite{Levoy:1996:LFR:237170.237199}.

Stereo 3D cinematography addresses a small part of this issue:  humans have two eyes and each provides visual information from its respective point of view. Stereo imagery therefore provides two different views, providing the disparity cue, but, as the human eye is not a pinhole device, the missing visual cues need to be approximated, creating restrictions for the imagery that can be displayed and problems for the audience ranging from discomfort up to headaches.

Ultimately, we will need plenoptic displays to achieve True AR, and here, technology is in its infancy, although current research prototypes\cite{lanman} provide limited accommodation cues for single users, they fall short in other domains, mainly resolution, and there is still a lot of ground to cover before they scale up to multiple users. 

There are two fundamental ways to embed light fields of virtual objects into reality.  First, we can inject these light fields into the user's direct view. Second, we can capture the light field of the user's environment, fuse it with the light field that we want to embed, and display this combination to the user.  Each approach has its advantages and disadvantages. The second approach is technically much more difficult, but has inherent advantages, including the correct handling of occlusions between real and virtual objects. In the next section, we discuss light field capture in more detail.

\subsection{Capturing Light Fields}

Just as a projector is the dual of a camera, in the sense that their geometrical configurations are the same and that the former emits light that the other captures, plenoptic sensors and displays also have much in common.  However, dense light field recording and display is still a high-dimensional, difficult problem, and thus approaches need to be employed that are built around asymmetric assumptions. As we will argue in this section, capture is actually easier to implement than display.

A plenoptic sensing device is usually constructed with simplifying assumptions in mind regarding the environment structure, such as restricting the scene to Lambertian surfaces (ones that look equally bright from every viewing configuration) or assuming occlusion-free scene geometry. In those cases, the scene around the viewer has an underlying model with fewer dimensions (possibly, merely a colored depth map from a certain point of view), and powerful 3D reconstruction algorithms may be able to generate a sufficiently complete scene description from only two input views. In the sense of sampling theory, this corresponds to observing the plenoptic function with point sampling from two possibly widely spaced points.

Again in the sense of sampling theory, a display device reconstructs a continuous light field to be observed by a human observer. Even if a renderable representation is known with arbitrary precision, a display needs to reconstruct a full, continuous function to be observed by the HVS.  An advantage of display technology is that the HVS is forgiving in some ways, and generating a visually equivalent light field signal---one the HVS could not distinguish from the perfect signal---is sufficient. Color perception, for instance, can be fully achieved with few primary colors in narrow spectral bands; even three primaries in red, green, and blue go a long way towards covering a visually complete color gamut.  However, this is only possible because the spectral sensitivity of the HVS contains wide low-pass prefilter kernels in the sampling sense.

For spatial signal variations---different view directions---dense sampling usually solves the problem; retina displays and 4K televisions come increasingly close to the resolution limits of the HVS. The extended aperture pupil of the human eye also acts in the fashion of a low-pass filter on the incident light distribution---however, with varying pupil diameter and varying refractive power in the lenses, a coarsely sampled input is insufficient, and, as many samples (images for different viewpoints) have to be generated for each pupil in order to achieve visually convincing impressions after filtering with the aperture, a light field display needs orders of magnitude more individually controllable basis elements for reconstruction than a light field sensor would need for sampling.

Hence---the unpredictability of technical progress notwithstanding---we would expect that light field sensors would stay ahead of light field displays for the foreseeable future.

\section{Facing Ethics: Non-Technical Barriers} 
What are the ethical and societal implications of True AR? Since True AR is a new technology, the question of power, in the Marxist sense, is inevitable: who will control its deployment, who will ultimately be in control of its augmented content, and for what purposes will it be used?  Will individuals be freed, or, on the contrary, be locked into purely commerce-driven experiences?
\begin{marginfigure}%
  \includegraphics[width=\linewidth]{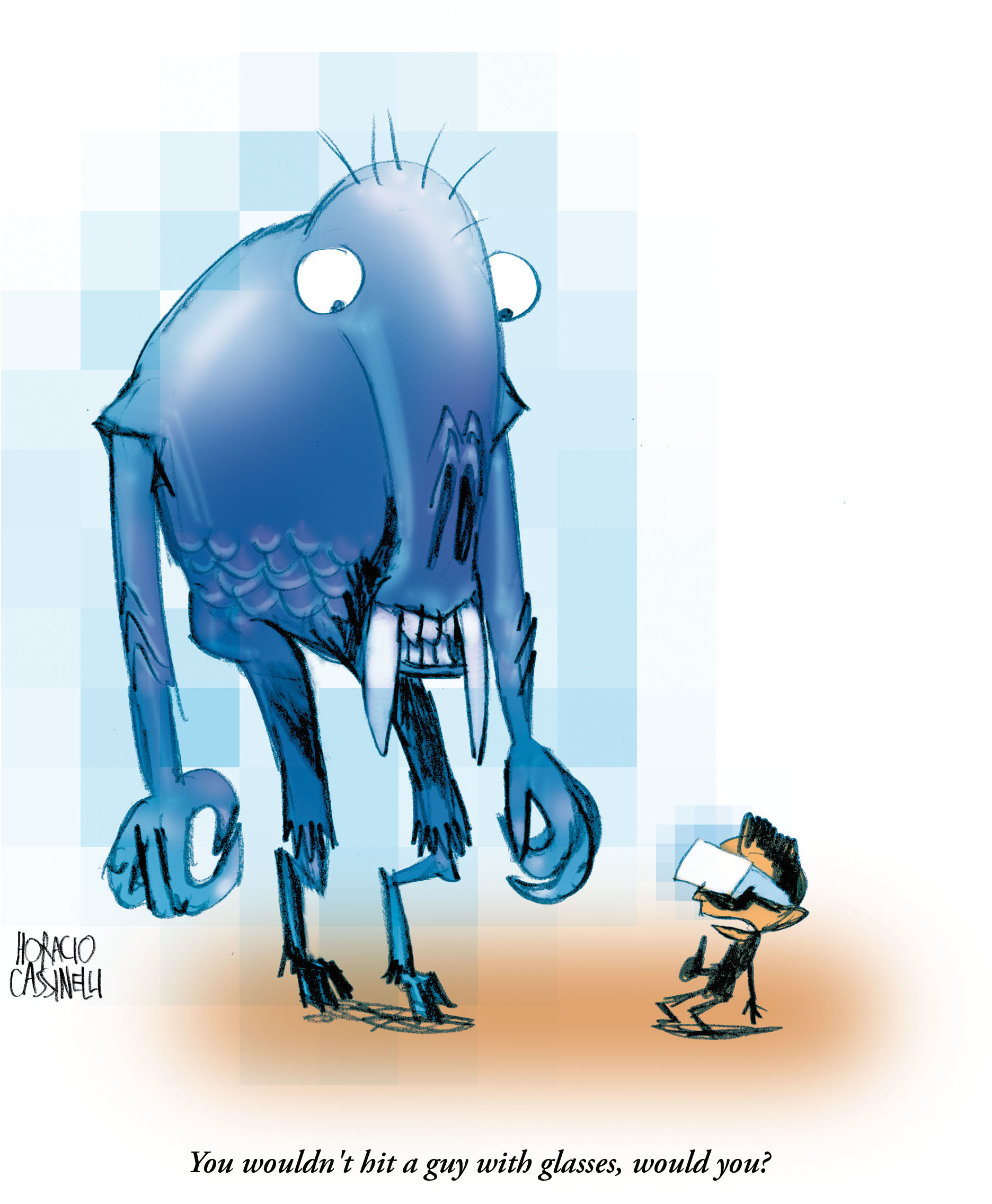}
  \caption{Used with permission of Horacio Cassinelli.}
  \label{fig:marginfig}
\end{marginfigure}

These questions, of course, address the heart of any technology representing and manipulating information, from books and the Gutenberg revolution up until the recent scandals around NSA mass surveillance of electronic communications.  However, we believe that True AR, by purposefully fusing information with the real world and dissolving any remain of discriminatory doubt, will be the most powerful medium that humanity ever had at its disposal. We will no longer observe the shadows in Plato's Cave, out of which only philosophers can glance a sight of reality; but, a ubiquitously interwoven stream of shadows and reality. Based on purposefully manipulated illusions, we will be incapable of telling what is there from that which others, Artificial Intelligences or humans, may want us to believe is there.

Equally interesting, and equally worrisome, is the direct impact that True AR will have on everyday life at the level of the individual: Will it enhance quality of life and human-to-human communication by promoting a deeper understanding of the true reality by letting us see past the world of illusion and deceit; for example, by informing us of the origin of a product or the way it was manufactured? Or, will it be the very instrument that will isolate and project us into a world of delusion?  For Buddhists, reality is a dream and the Samsara is filled with stimuli that prevent us from awakening, because we do not pause to meditate, but we react to them. Reacting, instead of being in control, is one of the worst consequences of misused information technology; for example, several studies have shown how the continuous use of social media has been decreasing attention spans among teenagers\cite{ophir}. In this sense, True AR may add layers to the illusion, or can, on the contrary, help us to see through them and filter the unimportant from the transcendent.

Let's get back to our first question, the question of power, which is inherent to any medium. Free access to information and public schooling is a good measure of social equality. Similarly, a 2011 report for the United Nations Human Right Council\cite{rue} led the media to suggest that Internet access is a fundamental human right. However, True AR presents a peculiar characteristic: it has the power to produce a radical divide of humanity into those who will live in an augmented world and others who will dwell in a world much less rich in information. This divide will affect the opportunities of individuals more profoundly than any previous technology, as it can continuously affect the perception of our immediate physical environment. A precursor of the upcoming debates was the discussion around Google Glass users in the San Francisco area, sometimes referred to as ``Glassholes.''  Socially minded individuals raised questions about the powerful discriminating effect that such a public display of this early form of transhumanism has.  We could observe a visceral, democratic reaction to an early form of the upcoming debates that ensue about True AR users in the future.

Our second question relates to the possible misdirection of users. Let us now ignore the bias that the establishment, the market, or whatever other social force inevitably introduces into any experience. The question could then be rephrased as follows: what will happen when True AR will confer infinite depth of perception into our everyday life? Indeed, one has to consider that this technology will not only be capable of deceiving our senses with perfectly physically co-located ``virtual'' objects, but will also augment our cognitive capabilities by prompting us with more abstract, but no less relevant data in real time. This will lead to an augmented self-awareness and ultimately a powerful meta-perception.
While interacting with others, we will be able to see through them, possibly into their personal and professional secrets; and, returning to our first question, if others cannot see us in the same way, a decent symmetry is being violated, resembling the asymmetry in surveillance practices that led Steve Mann to coin the concept of Sousveillance\cite{mann}. For example, using AR to see through walls has already been demonstrated\cite{xray}; rapid 3D modeling using Google's Project Tango could make this capability become a ubiquitous reality. AR Firewalls could address this problem; another way would be to always enforce symmetric AR; or, we could use the very same AR technologies to signal the presence of another AR user and invisible layers to one another. The question of power is closely related to the symmetry of AR, both form the technological point of view, as well as the ethical point of view.

Will True AR become a democratic, symmetric technology?  And if so, how far into the future will it take us?  Finally, if this happens, how will this total information transparency affect our everyday relations?   Information, including affective information, works as a currency between human beings. Whether it is natural or artificial, its scarcity makes it valuable and helps us trade in the world of personal relations. Not being able to hide anything may have the effect of devaluing interpersonal relations, but also erasing differences and avoiding the abuse of power.

\section{Conclusions} 
As our discussions at the symposium and in this article turn to an end, we are left with two main observations. On the one hand, we face the insight that the realization of True AR has to be built on simultaneous achievements that transcend the boundaries between disciplines traditionally thought not to be related to AR at all:  in addition to optics, computer graphics, and computer vision; later on, we will require input from disciplines including ethics, art, philosophy, and the social sciences to guide its applications.

On the other hand, the continuing technological progress that drives us ever closer to True AR requires us to define our goal posts more clearly. One possible approach for this could be an AR Turing Test, as a protocol to objectively determine whether True AR has been achieved. Therefore, we strongly believe that the development of an AR Turing Test must be added to the AR research agenda. As our discussions have shown, it is not straightforward to define such a test. Nevertheless, it is crucial, in order to have a measureable goal for the attempts of others and ourselves to erase the boundary between real and virtual.

\section*{Acknowledgements} 
We thank the Mirai-Kaitaku Colloquium of Nara Institute of Science and Technology (NAIST) for funding the symposium that led to this article. We are grateful to Hirokazu Kato and Goshiro Yamamoto (both NAIST) for their support in arranging this symposium. Last, but not least, we would like to thank all participants at the symposium. 

\begin{marginfigure}
  \includegraphics{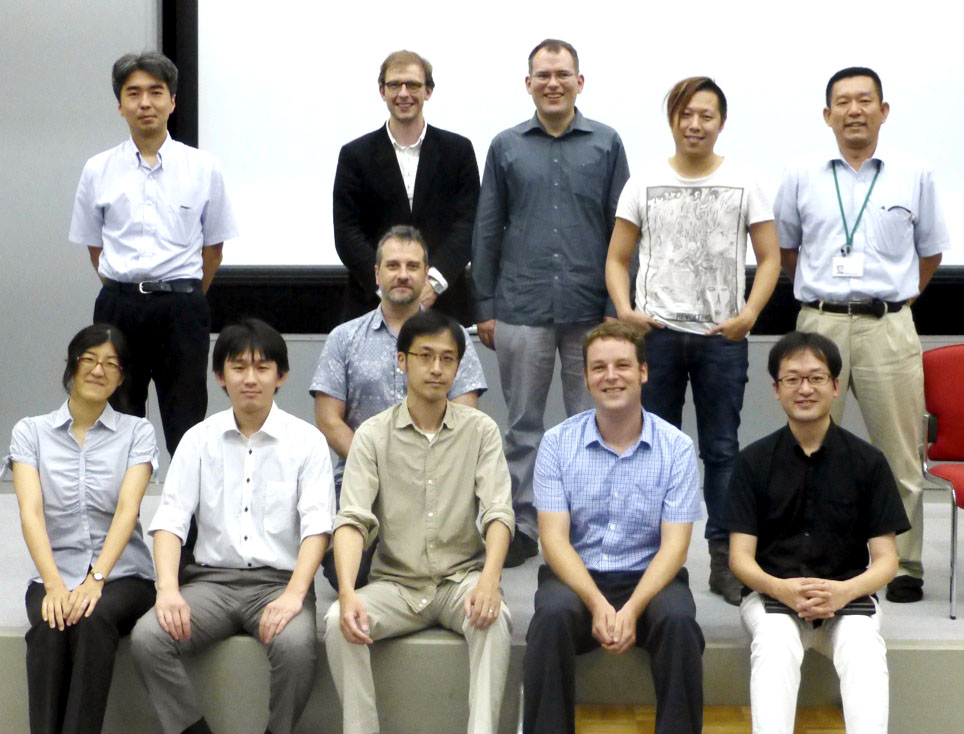}
  \caption{Group photo of symposium speakers from left to right then back to
  front: Tomokazu Sato (NAIST), Richard Newcombe (University of Washington),
  Martin Fuchs (Universit\"at Stuttgart), Hao Li (University of Southern California), Masayuki Kanbara (NAIST), Alvaro Cassinelli (University of Tokyo), Mai Otsuki (Tsukuba University), Goshiro Yamamoto (NAIST), Takeo Igarashi (University of Tokyo), Christian Sandor (NAIST), and Masahiko Inami (Keio University).}
  \setfloatalignment{b}
\end{marginfigure}

\bibliography{tufte-style.bib}
\bibliographystyle{plainnat}
\end{document}